\documentstyle[aps,prb,epsf]{revtex}

\newcommand{\pdag}{{\phantom{\dagger}}}
\newcommand{\bq}{\begin{equation}}
\newcommand{\eq}{\end{equation}}
\newcommand{\bn}{\begin{eqnarray}}
\newcommand{\en}{\end{eqnarray}}

\begin{document}
\draft
\title{Effect of the Kondo correlation on thermopower in a Quantum Dot}
\author{Bing Dong and X. L. Lei}
\address{Department of Physics, Shanghai Jiaotong University, 1954 Huashan Road, Shanghai 
200030, P. R. China}
\maketitle
\begin{abstract}
In this paper we study the thermopower of a quantum dot connected to two leads in the 
presence of Kondo correlation by employing a modified second-order perturbation scheme at 
nonequilibrium. A simple scheme, Ng's ansatz [Phys. Rev. Lett. {\bf 76}, 487 (1996)], is 
adopted to calculate nonequilibrium distribution Green's function and its validity is 
further checked with regard to the Onsager relation. Numerical results demonstrate that 
the sign of the thermopower can be changed by tuning the energy level of the quantum dot, 
leading to a oscillatory behavior with a suppressed magnitude due to the Kondo effect. We 
also calculate the thermal conductance of the system, and find that the Wiedemann-Franz 
law is obeyed at low temperature but violated with increasing temperature, corresponding 
to emerging and quenching of the Kondo effect.        

\end{abstract}

\section{Introduction}
Observation of the Kondo effect in a semiconductor quantum dot (QD) 
\cite{Goldhaber,Cronenwett} which provided a testing ground of the quantum behavior of 
electron wave functions and many-body effects, has stimulated a great deal of experimental 
\cite{Goldhaber1,Sasaki} and theoretical \cite{Meir1,Meir2,Yeyati,Craco,Ng} investigation. 
So far, many interesting features in QD, such as Kondo-assisted enhancement of 
conductance, its specific temperature dependence, a peak splitting in a magnetic field, 
zero-bias maximum of differential conductance in the Kondo regime, and singlet-triplet 
Kondo effect have been successfully explored in early theoretical works, which described 
the electron transport through QD using the well-known impurity Anderson model. 
\cite{Anderson} 
 
A convenient way of describing transport through an Anderson impurity under 
out-of-equilibrium condition is to employ the nonequilibrium Green's function (GF) 
technique of Keldysh. In the framework of nonequilibrium GF, one needs a consistent 
calculations of retarded (advanced) and distribution (lesser) Greens's functions in order 
to study the out-of-equilibrium properties, while only the retarded GF is needed when 
studying equilibrium properties of the system. In literature, the single-impurity Anderson 
model has been extensively studied for more than thirty years and many numerical and 
analytical methods have been developed to explore its equilibrium properties, such as 
equation-of-motion (EOM) combined with the decoupling approximation, \cite{Lacroix} the 
second-order perturbation theory (SOPT) for on-site Coulomb interaction $U$, \cite{YYHZ} 
slave-boson non-crossing approximation (NCA). \cite{Coleman,Bickers} However, it is still 
a great challenge to evaluate the distribution GF of the strongly correlated systems under 
out-of-equilibrium condition, because one has to deal with both the on-site Coulomb 
interaction and the tunnelings between the dot and reservoirs simultaneously. To our 
knowledge, there were three attempts to surmount this technical difficulty. The first one 
has been carried out by Meir, \cite{Meir2} who generalized the NCA to the nonequilibrium 
situation basing on the Coleman's slave-boson method in the limit of infinite Coulomb 
interaction $U\rightarrow\infty$. \cite{Coleman} This scheme transforms the strong 
correlation Hamiltonian into a noninteracting equivalent one by introducing several 
auxiliary boson operators. Therefore, it evades the dilemma how both the Coulomb 
interaction $U$ and the mixing between QD and the two leads can be treated simultaneously 
in deriving the retarded and lesser (greater) self-energies of QD. It is well known that 
NCA provides a good description for the investigation of the excitation spectra of QD, but 
the necessary Fermi liquid behavior was not reproduced at low energy and low temperature 
limit. To extend the slave-boson method to study Kondo-type transport through double QDs 
at low temperature, espatially at zero temperature, a slave-boson mean field approach has 
been presented by replacing the slave-boson operators with their expectation values. 
\cite{Aono} Very recently, a finite-$U$ slave-boson mean field scheme has been developed 
by us to explore Kondo-type transport through QD \cite{Dong1} and double QDs \cite{Dong2}. 
Nevertheless, these slave-boson mean field approaches are at the same level of 
approximation in deriving the interacting retarded and lesser self-energies. 

The third attempt has been made by Ng, \cite{Ng} who has developed an approximative 
scheme, i.e., Ng's ansatz, to obtain the lesser GF from EOM. It is impossible to obtain 
straightforwardly the distribution GF from EOM without introducing additional assumptions. 
Ng assumed that the interacting greater or lesser self-energy of QD is related to 
noninteracting one, and the ratio parameter is determined by the interacting retarded 
self-energies derived by means of EOM. This simple and effective ansatz has later been 
applied in more complicated structures involved QD, such as normal metal-QD-superconductor 
\cite{Fazio}, and superconductor-QD-superconductor \cite{Cho}. Nevertheless, except for 
these three advantages pointed out by Ng in Ref.\onlinecite{Ng}, up to now, no further 
inspectation about the validity of the ansatz has been presented.

In this paper we will provide another validity check of Ng's ansatz by analyzing the 
particle current and thermal flux through QD connecting to two reservoirs on the basis of 
Anderson single-impurity model by means of nonequilibrium GF with the help of Langreth 
continuation rules and Ng's ansatz. We find that the resulting particle current and heat 
flux driven by bias voltages and temperature gradients between two leads satisfy the 
Onsager relation in nearly equilibrium regime, \cite{Onsager} thus provide a natural check 
on the validity of derivation of the lesser GF.

To date, most theoretical calculation and experimental measurement on the thermopower of 
QD have focused in the Coulomb blockade (CB) regime, \cite{Andreev} which have revealed 
that when the gate voltage defining QD is swept, the thermopower oscillates about zero 
(sawtooth behavior) with a period equal to that of the CB oscillations in electric 
conductance. However, thermopower across QD in the Kondo regime is still much less 
studied. \cite{Kim,Boese} The second purpose of this paper is to investigate the Kondo 
effect on the linear thermopower of QD basing on Ng's ansatz. 

We organize the rest parts of the paper as following. In the second section we derive 
particle current and thermal flux formulas through interacting QD within the framework of 
nonequilibrium GF. The electric and thermoelectric transport coefficients are derived in 
the presence of both chemical potential and temperature gradients between two leads and 
automatically satisfy Onsager relation in the linear transport regime. We note that within 
this approximation, the same current formula as that in Ref.\onlinecite{Meir3} is derived 
without the presumption of proportional coupling $\Gamma_{L}(\omega)=\lambda 
\Gamma_{R}(\omega)$. In section 3, numerical calculation of linear Kondo-type thermopower 
$S$ in QD are reported as a function of gate voltage and temperature, which shows a 
gate-voltage-controlled change of sign and largely suppressed oscillatory magnitude due to 
the Kondo effect. We also discuss the thermal conduction coefficient $\kappa$ and its 
violations of the well-known Wiedemann-Franz law. Finally, a conclusion is given in 
Section 4.     

\section{Thermal Current Formula and Onsager Relation}

Transport through a QD coupled to two reservoirs in the presence of external voltages and 
temperature gradient between two reservoirs can be described by the Anderson single 
impurity model:
\bq
H=\sum_{\eta, k, \sigma}\epsilon _{\eta k \sigma}^{\pdag}c_{\eta k \sigma }^{\dagger 
}c_{\eta k \sigma }^{\pdag} + \epsilon_{d} \sum_{\sigma} c_{d\sigma }^{\dagger} c_{d\sigma 
}^{\pdag} + Un_{d\uparrow }n_{d\downarrow }
+ \sum_{\eta, k, \sigma} (V_{\eta} c_{\eta k \sigma }^{\dagger} c_{d\sigma }^{\pdag} + 
{\rm {H.c.}}),
\label{hamiltonian1}
\eq
where $\epsilon _{\eta k \sigma }$ represents the conduction electron energy of the lead 
$\eta$, and $\epsilon_{d}$ is the discrete energy level in the QD. $c_{\eta k 
\sigma}^{\dagger }$ ($c_{\eta k \sigma }$) are the creation (annihilation) operators for 
electrons in the lead $\eta$ ($={\rm L,R}$), while $c_{d\sigma}^{\dagger}$ ($c_{d\sigma}$) 
for electrons in the QD. When a total external voltage $V$ is applied between the two 
leads, their chemical potential difference is $\mu_{\rm L}-\mu_{\rm R}=eV$. The two leads 
are assumed to be in local equilibrium with respective temperature $T_{\eta}$ and their 
distribution functions are given by $f_{\eta}(\omega)=[1+\exp{(\omega-\mu_\eta)/k_{\rm 
B}T_{\eta}}]^{-1}$ ($\eta=$L or R). This assumption is practically correct because the two 
reservoirs respond to an external perturbation much faster than the center region, {\it 
i.e.}, the QD. The other parameters $U$ and $V_{\eta}$, stand for the Coulomb interaction, 
and the coupling between the QD and the reservoir $\eta$, respectively. For the transport 
problem concerned in this paper, electrons in the QD are in a nonequilibrium state, to be 
determined by their coupling to two leads and by the applied voltage. In order to describe 
the nonequilibrium state of electrons, we define the retarded (advanced) and lesser 
(greater) GFs for the QD as follows: $G_{d\sigma}^{r(a)}(t,t')\equiv \pm i\theta (\pm t 
\mp t')\langle \{ c_{d\sigma}^{\pdag} (t) , c_{d\sigma}^{\dagger}(t') \}\rangle$, 
$G_{d\sigma}^{<}(t,t')\equiv i\langle c_{d\sigma}^{\dagger}(t')c_{d\sigma}^{\pdag} (t) 
\rangle$ and $G_{d\sigma}^{>}(t,t')\equiv-i\langle c_{d\sigma}^{\pdag} (t) 
c_{d\sigma}^{\dagger}(t') \rangle$. 

The particle current $J_{\eta}$ and energy flux $J_{E\eta}$ flowing from the lead $\eta$ 
to the QD can be evaluated, respectively, from the rate of change of the electron number 
operator $N_{\eta}(t)=\sum_{k, \sigma} c_{\eta k\sigma}^{\dagger}(t) c_{\eta 
k\sigma}^{\pdag}(t)$ and the rate of change of energy operator $H_{\eta}(t)=\sum_{k, 
\sigma} \epsilon_{\eta k\sigma}^{\pdag}c_{\eta k\sigma}^{\dagger}(t) c_{\eta 
k\sigma}^{\pdag}(t)$ of the lead $\eta$: \cite{Mahan}
\begin{eqnarray}
{J}_{\eta}(t)&=&-\frac{1}{\hbar}\Big\langle \frac{d{N}_{\eta}}{dt}\Big\rangle=
-i\frac{1}{\hbar}\Big\langle \Big [H, \sum_{k,\sigma}c_{\eta k \sigma}^{\dagger}(t)
c_{\eta k \sigma}(t)\Big ] \Big\rangle=i\frac{1}{\hbar} \Big\langle \sum_{k,\sigma} 
[V_\eta^{\pdag}  c_{\eta k\sigma}^\dagger(t) c_{d\sigma}^{\pdag}(t)-V_\eta^{*} 
c_{d\sigma}^\dagger(t) c_{\eta k\sigma}^{\pdag}(t)] \Big\rangle, \label{i} \\
{J}_{E\eta}(t)&=&-\frac{1}{\hbar}\Big\langle \frac{d{H}_{\eta}}{dt}\Big\rangle=
-i\frac{1}{\hbar} \Big\langle \Big [H, \sum_{k, \sigma} \epsilon_{\eta k\sigma }^{\pdag} 
c_{\eta k\sigma}^{\dagger}(t) c_{\eta k \sigma}(t)\Big ] \Big\rangle 
=i\frac{1}{\hbar}\Big\langle \sum_{k,\sigma} \epsilon_{\eta k\sigma}^{\pdag} 
[V_\eta^{\pdag} c_{\eta k\sigma}^\dagger(t) c_{d\sigma}^{\pdag}(t)-V_\eta^{*} 
c_{d\sigma}^\dagger(t) c_{\eta k\sigma}^{\pdag}(t)] \Big\rangle, \label{e}
\end{eqnarray}
which involve the time-diagonal parts of the correlation functions: $G_{d\sigma,\eta k 
\sigma}^{<}(t,t')\equiv i\langle c_{\eta k\sigma}^{\dagger}(t')
c_{d\sigma}^{\pdag}(t)\rangle$ and $G_{\eta k\sigma,d\sigma}^{<}(t,t')\equiv i\langle 
c_{d\sigma}^{\dagger}(t')c_{\eta k\sigma}^{\pdag}(t)\rangle$. According to 
Ref.\,\onlinecite{Mahan}, the thermal flux $J_{Q\eta}$ flowing from the lead $\eta$ to the 
QD is determined as
\bq
J_{Q\eta}=J_{E\eta}-\mu_{\eta} J_{\eta}. \label{q}
\eq  
With the help of the Langreth analytic continuation rules, \cite{Langreth} we obtain the 
following expressions: 
\bn
J_{\eta}&=&i\int \frac{d\omega}{2\pi \hbar} \Gamma_{\eta}(\omega) \sum_{\sigma}  \left \{ 
G_{d\sigma}^{<}(\omega) + f_{\eta}(\omega) \left [ G_{d\sigma}^{r}(\omega) - 
G_{d\sigma}^{a}(\omega) \right ] \right \}, \label{ii} \\  
J_{Q\eta}&=&i\int \frac{d\omega}{2\pi \hbar} \Gamma_{\eta }(\omega) \sum_{\sigma}  
(\omega-\mu_{\eta})\left \{ G_{d\sigma}^{<}(\omega) + f_{\eta}(\omega) \left [ 
G_{d\sigma}^{r}(\omega) - G_{d\sigma}^{a}(\omega) \right ] \right \}, \label{qq}
\en
in terms of the QD's GFs in the Fourier space. Here $\Gamma_{\eta} (\omega)=2\pi \sum_{k, 
\sigma}|V_{\eta}|^2 \delta (\omega-\epsilon_{\eta k\sigma})$ denotes the strength of 
coupling between the QD level and the lead $\eta$.

In the presence of both a strong Coulomb interaction in QD and tunnelings between QD and 
leads, it is difficult to evaluate the retarded GF $G_{d\sigma}^{r}$ of QD accurately. 
Several approximation schemes have been proposed in literature to derive 
$G_{d\sigma}^{r}$, such as EOM with the decoupling approximation, SOPT, etc, where a 
retarded (advanced) self-energy for the interacting QD is written as 
$\Sigma^{r(a)}=\Sigma_{0}^{r(a)}+\Sigma_{i}^{r(a)}$, with $\Sigma_{0}^{r(a)}=\mp 
i\sum_{\eta}\Gamma_{\eta}(\omega)/2$ being the noninteracting part coming from the 
tunneling of electrons from the impurity state to outside leads, and $\Sigma_{i}^{r(a)}$ 
being the interacting part derived within these approximation approaches. The retarded 
(advanced) GF $G_{d\sigma}^{r(a)}$ thus has the form
\bq
G_{d\sigma}^{r(a)}(\omega)=\frac{1}{\omega-\epsilon_{d}-\Sigma^{r(a)}(\omega)}. \label{gr}
\eq
Unfortunately, one can not straightforwardly get the lesser GF for the strongly correlated 
systems under out-of-equilibrium circumstance, as does for the retarded GF. Several years 
ago, Ng proposed a simple scheme to obtain the lesser GF $G_{d\sigma}^{<}$ from the 
retarded and advanced terms in order to study ac Kondo resonances in nonlinear transport 
basing on EOM approach. He assumed that $\Sigma^{<}(\omega)=A\Sigma_{0}^{<}(\omega)$ and 
$\Sigma^{>}(\omega)=A\Sigma_{0}^{>}(\omega)$ where $A$ is an unknown function, while 
$\Sigma_{0}^{<}=i\sum_{\eta} \Gamma_{\eta}(\omega) f_{\eta}(\omega)$ and 
$\Sigma_{0}^{>}=-i\sum_{\eta}\Gamma_{\eta}(\omega) [1-f_{\eta}(\omega)]$ are 
noninteracting lesser and greater self-energies. These lesser and greater self-energies 
are requested to satisfy the Keldysh relation 
$\Sigma^{<}-\Sigma^{>}=\Sigma^{r}-\Sigma^{a}$, leading to
\bq
A=\frac{\Sigma^{r}-\Sigma^{a}}{\Sigma_{0}^{r}-\Sigma_{0}^{a}},
\eq
or in explicit form,
\bq
\Sigma^{<}(\omega)=-2i\frac{\sum_{\eta} \Gamma_{\eta}(\omega) f_{\eta}(\omega)} 
{\sum_{\eta} \Gamma_{\eta}(\omega)} {\rm Im} \Sigma^{r}. \label{Ngseless}
\eq
The lesser GF $G_{d\sigma}^{<}=\Sigma^{<}|G_{d\sigma}^{r}|^2$ is thus obtained. This is 
the central result of Ng's scheme. It has three advantages, initially addressed by Ng, 
that (i) it is exact in the equilibrium limit $\mu_{L}=\mu_{R}$, (ii) it is exact in the 
noninteracting ($U=0$) limit under general nonequilibrium situations, and (iii) the 
continuity equation $J_{L}(t)=-J_{R}(t)$ is automatically satisfied in the steady state 
limit. \cite{Ng} With the help of this ansatz, as long as one obtains the retarded GF 
properly describing the strongly correlated system from certain approximative method, the 
lesser GF can be derived, and thus the transport problem can be investigated. By means of 
Eq.\,(\ref{Ngseless}), we easily obtain
\bn
J_{\eta}&=&-\frac{2}{h}\int d\omega \Gamma(\omega) \left [ f_{\eta}(\omega) - 
f_{\bar{\eta}}(\omega)  \right ] {\rm Im}G_{d\sigma}^{r}(\omega), \label{iii} \\  
J_{Q\eta}&=&-\frac{2}{h}\int d\omega \Gamma(\omega) (\omega - \mu_{\eta})\left [ 
f_{\eta}(\omega) - f_{\bar{\eta}}(\omega) \right ] {\rm Im}G_{d\sigma}^{r}(\omega), 
\label{qqq}
\en
where $\Gamma(\omega)\equiv \Gamma_{L}(\omega) \Gamma_{R}(\omega)/[\Gamma_{L}(\omega) + 
\Gamma_{R}(\omega)]$ and $\bar{\eta}\neq \eta$. Note that here we arrive at exactly the 
same current formula (\ref{iii}) as that in Ref.\onlinecite{Meir3} without introducing the 
assumption of a proportional coupling $\Gamma_{L}(\omega)=\lambda \Gamma_{R}(\omega)$ 
($\lambda$ is a constant). Since the pioneer work of Ng, \cite{Ng} several attempts have 
been made to generalize this ansatz to study Kondo-type transport in more complicated 
devices containing interacting QD, such as normal metal-QD-superconductor \cite{Fazio} and 
superconductor-QD-superconductor \cite{Cho}. Nevertheless, the validity of this ansatz is 
worth being further examined. As mentioned in the introduction, verification of the 
Onsager relation is a natural choice for this purpose.

The Onsager relation is concerned with the linear response of the particle current 
$J_{\eta}$ [Eq.\,(\ref{iii})] and the heat flux $J_{Q\eta}$ [Eq.\,(\ref{qqq})] driven by 
small bias voltages $\mu_{\eta}-\mu_{\bar{\eta}}=\delta\mu$ and small temperature 
gradients $T_{\eta}-T_{\bar{\eta}}=\delta T$:
\bn
J_{\eta}&=& -{\cal L}_{11}\frac{\delta\mu}{T} - {\cal L}_{12} \frac{\delta T}{T^2} 
=-\frac{2}{h}\int d\omega {\cal T}(\omega) \left [ \left (\frac{\partial f_{\eta}(\omega) 
}{\partial\mu} \right )_{T} \delta\mu + \left( \frac{\partial f_{\eta}(\omega)}{\partial 
T}\right )_{\mu} \delta T \right ] , \label{iii-linear}\\
J_{Q\eta}&=& -{\cal L}_{21} \frac{\delta\mu}{T} - {\cal L}_{22} \frac{\delta T}{T^2} 
=-\frac{2}{h}\int d\omega {\cal T}(\omega) (\omega-\mu_{\eta}) \left [ \left 
(\frac{\partial f_{\eta}(\omega) }{\partial\mu} \right )_{T} \delta\mu + \left( 
\frac{\partial f_{\eta}(\omega)}{\partial T}\right )_{\mu} \delta T \right ], 
\label{qqq-linear}
\en
with ${\cal T}(\omega)=\Gamma(\omega)\, {\rm Im} G_{d\sigma}^{r} 
(\omega)_{\delta\mu=0,\delta T=0}$ and
\bn
{\cal L}_{11}&=&\frac{2T}{h}\int d\omega {\cal T}(\omega) \left (\frac{\partial 
f_{\eta}(\omega) }{\partial\mu} \right )_{T},\; \; \; \; {\cal L}_{12}=\frac{2T^2}{h}\int 
d\omega {\cal T}(\omega) \left( \frac{\partial f_{\eta}(\omega)}{\partial T}\right 
)_{\mu}, \label{l11} \\
{\cal L}_{21}&=&\frac{2T}{h}\int d\omega {\cal T}(\omega) (\omega-\mu_{\eta}) \left ( 
\frac{\partial f_{\eta}(\omega) }{\partial\mu} \right )_{T},\; \; \; \; {\cal 
L}_{22}=\frac{2T^2}{h}\int d\omega {\cal T}(\omega) (\omega-\mu_{\eta}) \left ( 
\frac{\partial f_{\eta}(\omega)}{\partial T}\right )_{\mu}. \label{l22}
\en
We can easily see that Ng's ansatz Eq.\,(\ref{Ngseless}) preserves Onsager relation ${\cal 
L}_{12}={\cal L}_{21}$ automatically. Furthermore, the result is independent of the 
approximation adopted in deriving the retarded GF.
 
\section{Thermoelectric Effects in the Presence of Kondo Correlation}

As shown in Eq.\,(\ref{iii-linear}), both the bias voltage and the temperature gradient 
between two reservoirs can give rise to particle current. The current induced purely by a 
small bias voltage reflects the electric conductance $G=-\frac{e^2}{T}{\cal L}_{11}$, 
while the thermopower $S$ measures the voltage difference needed to eliminate the current 
due to the temperature gradient between the leads, given in linear regime by 
\bq
S=-\frac{1}{eT}\frac{{\cal L}_{12}}{{\cal L}_{11}}.
\eq
In this situation, we can simplify the thermal flux Eq.\,(\ref{qqq-linear}) as 
$J_{Q\eta}=-\kappa \delta T$, in which 
\bq
\kappa=\frac{1}{T^2}\left ({\cal L}_{22}-\frac{{\cal L}_{12}^2}{{\cal L}_{11}}\right )
\eq
is the thermal conductance.      

Comparing the explicit expressions of ${\cal L}_{11}$ and ${\cal L}_{12}$ 
[Eq.\,(\ref{l11})], we can roughly address that in low temperatures, the electric 
conductance $G$ is determined by the transmission probability, or the density-of-state 
(DOS) of QD, at the Fermi energy of the leads, while the linear thermopower $S$ depends 
sensitively on the energy dependence of DOS, implying that it contains information 
different from the electric conductance. So far, the thermopower of QD in the kondo regime 
is still much less studied in literature. In this section, we attempt to numerically 
calculate the linear Kondo-type thermoelectric effects in QD. The remaining task is to 
choose a suitable approximative scheme, which can provide the retarded self-energy [or DOS 
$\rho(\omega)$] capable of properly describing the Kondo physics in a wide range of 
parameters of interacting QD. In the present paper, a modified SOPT developed in 
Ref.\,\onlinecite{Craco} is adopted, in which
\bq
\Sigma_{i}^{r}(\omega)=Un + \frac{a\Sigma^{(2)}(\omega)}{1-b\Sigma^{(2)}(\omega)}, 
\label{sopt}
\eq
with $a=n(1-n)/n_{0}(1-n_{0})$ and $b=(1-2n)/n_{0}(1-n_{0})U$. $n$ is the occupation 
number of the QD level which should be determined self-consistently. $\Sigma^{(2)}$ and 
$n_{0}$ are, respectively, the second order self-energy in $U$ and a fictitious particle 
number, both of which are obtained from the bare GF 
$G_{0}^{r}=1/(\omega-\epsilon_{d}-Un-\Sigma_{0}^{r})$.      

In actual calculation, an assumption that the tunneling strength is independent of 
incident energy is taken in the wide-band limit and a symmetric system 
$\Gamma_L(\omega)=\Gamma_R(\omega)=\Gamma$ is focused. In the following we take the 
coupling strength $\Gamma$ as the energy unit and the Fermi level of the lead to be zero.

In Figs.\,1(a)-(d) we plot the equilibrium DOS $\rho(\omega)$ for the QD with $U=7$ and 
several different energy levels $\epsilon_d=-2$, $-3.5$, $-5$ and $2$ at different 
temperatures. We can observe that the Kondo resonance peak in DOS is clearly resolved for 
the Kondo systems $\epsilon_d=-2$, $-3.5$ and $-5$ at low temperature $T=0.01$, but nearly 
dispeares at high temperature $T=1$. Of course there is no Kondo peak in DOS for the 
non-Kondo system $\epsilon_d=2$. In Fig.\,2, the calculated electric conductance $G$ (a), 
thermopower $S$ (b) and thermal conductance $\kappa$ (c) are displayed as functions of the 
gate voltage, i.e., the energy level in QD. As expected, the conductance demonstrates a 
single peak structure and nearly unitary limit near symmetric point $\epsilon_d=-U/2$ at 
very low temperature, splitting peaks with increasing temperature and a minimum at 
$\epsilon_d=-U/2$ for very high temperature. It manifests the equilibrium DOS of QD at the 
Fermi energy of the lead, $\rho(0)$, at low temperature, but not its concrete shape of the 
Kondo peak: symmetric or nonsymmetric (right or left of center) around the Fermi energy of 
the lead. In order to detect DOS in the whole range of energy, one has made an attempt to 
measure the differential conductance of QD. However, the finite bias voltages between two 
leads result in a large change of DOS, giving rise to a splitting of the Kondo peak. As 
mentioned above, an alternative way of addressing this problem is to explore the 
thermopower, because it is relevant with the product of the incident electron energy 
$\omega$ and the DOS of QD $\rho(\omega)$. For example, the perfectly symmetric shape of 
DOS around $\omega=0$ at the symmetric case $\epsilon_d=-3.5$, as shown in Fig.\,1(b), 
results in exactly zero of the thermopower; while the positive (negative) thermopower is 
attributed to slightly slanting of DOS towards the left (right) for $\epsilon_d=-2$ ($-5$) 
[Figs.\,1(a) and (c)], implying an oscillatory behavior around zero which can be 
controlled by the gate voltage. We also observe from Fig.\,2(b) that the magnitude of 
oscillating in thermopower is largely decreased with lowering temperature due to the 
Kondo-suppressed deviation of DOS from the symmetric shape. Fig.\,2(c) reveals that the 
thermal conductance $\kappa$ has similar behavior as the electric conductance $G$ at low 
temperature, but quite different at high temperature $T=1$. 

Fig.\,3 displays the temperature dependence of these three quantities. We observe from 
Fig.\,3(b) that the thermopower shows a logarithmic riseup with increasing temperature, 
leading to a broad maximum, and subsequently decreases in the high temperature regime. But 
temperature increasing can not cause a change of the sign of the thermopower, except for 
very high temperature where the Kondo effect is quenched. Fig.\,3(c) shows that the 
thermal conductance has similar temperature dependence for these different systems. As a 
result, the thermal conductance is not a suitable tool to explore the Kondo effect. The 
interesting physical quantity is the ratio between the thermal and electric conductance. 
The classical theory yields that thermal and electric transport in bulk metals satisfy the 
Wiedemann-Franz law $\kappa/TG=\pi^2/3e^2$. In mesoscopic system transport occurs through 
a small confined region, and consequently does not satisfy the Wiedemann-Franz law in 
general. We depict this ratio in Fig.\,4. Surprising recovery of the Wiedemann-Franz law 
is observed at very low temperature, where transport through QD is dominated by Kondo 
correlation, implying that the Landau Fermi liquid state is rebuilt in this situation. But 
substantial deviation from the classical value emerges with increasing temperature due to 
disappearance of the Kondo effect. This temperature behavior provides a good trademark for 
exploring the onset of Kondo correlation.       

\section{Conclusion}

We have studied the particle current and thermal flux through interacting QD on the basis 
of the nonequilibrium Green's function approach and Ng's ansatz. The advantage of the 
ansatz is its capability of evaluating the lesser (greater) GF from the retarded and 
advanced GFs derived in certain rational approximation scheme. The validity about the 
ansatz and the resulting linear transport coefficients have been examined in term of 
Onsager relation. We have also emphasized that the same electric current formula as in 
Ref.\onlinecite{Meir3} can be obtained without the assumption of proportional coupling 
$\Gamma_{L}(\omega)=\lambda \Gamma_{R}(\omega)$.      

In the wide band limit, a modified SOPT has been employed to calculate the retarded GF for 
interacting QD and with this retarded GF the thermoelectric effect in the Kondo regime has 
been investigated. We have found that the thermopower exhibits a oscillatory behavior 
around zero due to nonsymmetric shape of the Kondo peak in the DOS, giving rise to a 
change of sign of the thermopower which is controllable by tuning the gate voltage. These 
results demonstrate that measuring thermopower can provide useful information of the DOS 
in QD. Furthermore, our calculation reveals that the magnitude of oscillation is largely 
suppressed by the Kondo effect. Finally, we have explored the temperature characteristic 
of the thermoelectric effect and predicted that at low temperature regime thermal 
transport satisfies the classical Wiedemann-Franz law, which can be taken as a trace of 
the Kondo correlation.       


\section*{Acknowledgements}

This work was supported by the National Natural Science Foundation
of China, the Special Funds for Major State Basic Research 
Project (grant No. 2000683), the Ministry of Science and Technology of China, the
Shanghai Municipal Commission of Science and Technology, and the
Shanghai Foundation for Research and Development of Applied
Materials.


\section*{Figure Captions}

\begin{itemize}
\item[{\bf Fig.\,1}] The DOS of the interacting QD $U=7$ calculated from the modified SOPT 
with several energy levels $\epsilon_d=-2$ (a), $-3.5$ (b), $-5$ (c), $2$ (d) and 
different temperatures $T=0.01$, $0.5$, and $1$. In this and following figures, $\Gamma$ 
is chosen as energy unit.   

\vspace{1mm}

\item[{\bf Fig.\,2}] (a) The electric conductance $G$, (b) the thermopower $S$, and (c) 
the thermal conductance $\kappa$, as functions of the energy level for several different 
temperatures $T=0.01$, $0.5$, and $1$. The on-site Coulomb interaction in the QD is $U=7$.

\vspace{1mm}

\item[{\bf Fig.\,3}] (a) The electric conductance, (b) the thermopower, and (c) the 
thermal conductance vs temperature for the QD with $U=7$ and $\epsilon_d=-1$, $0$, and 
$-5$.

\vspace{1mm}

\item[{\bf Fig.\,4}] The ratio factor between the electric conductance and thermal 
conductance vs temperature for the QD with the same parameters as in Fig.\,3.

\end{itemize}

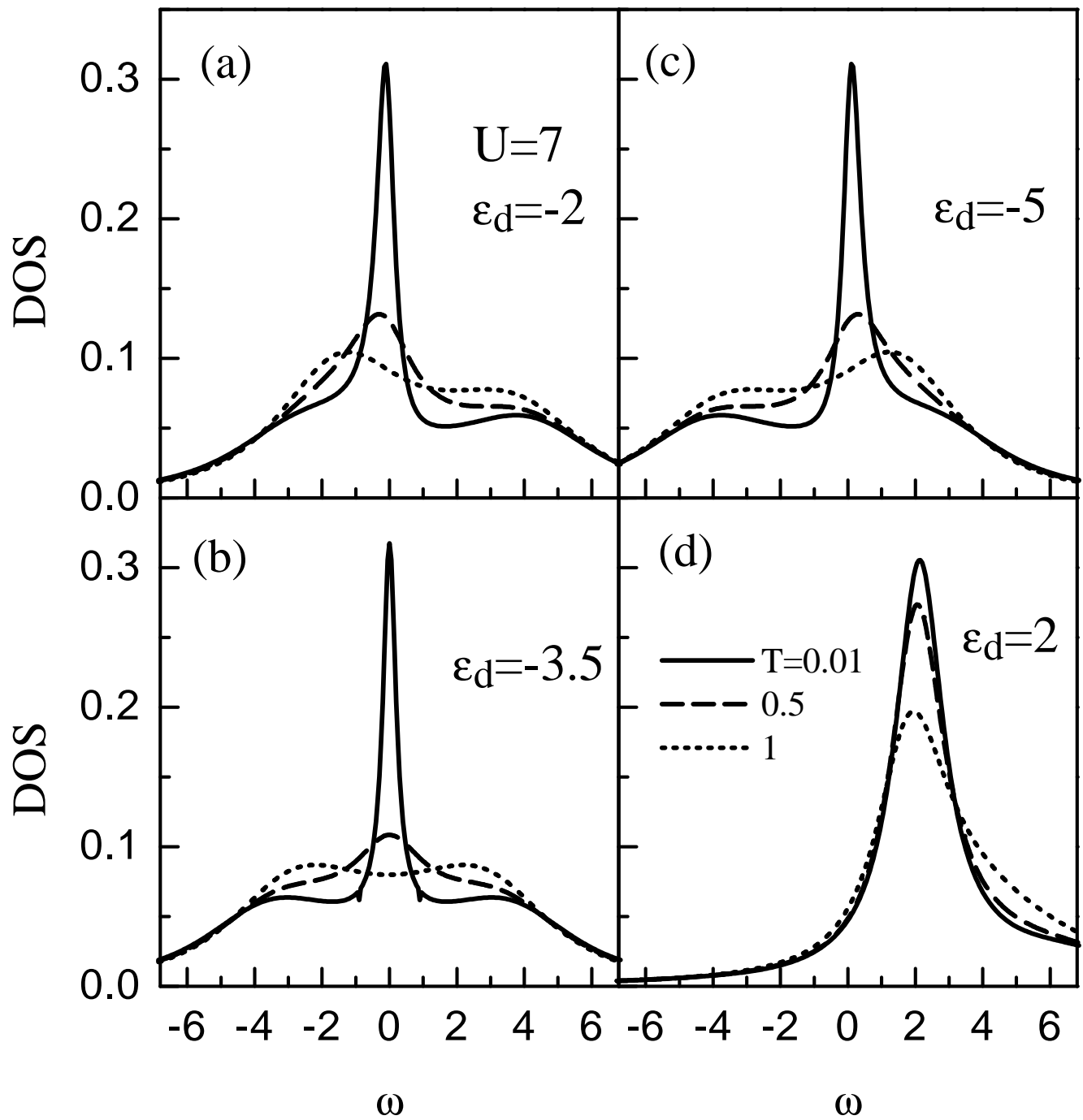
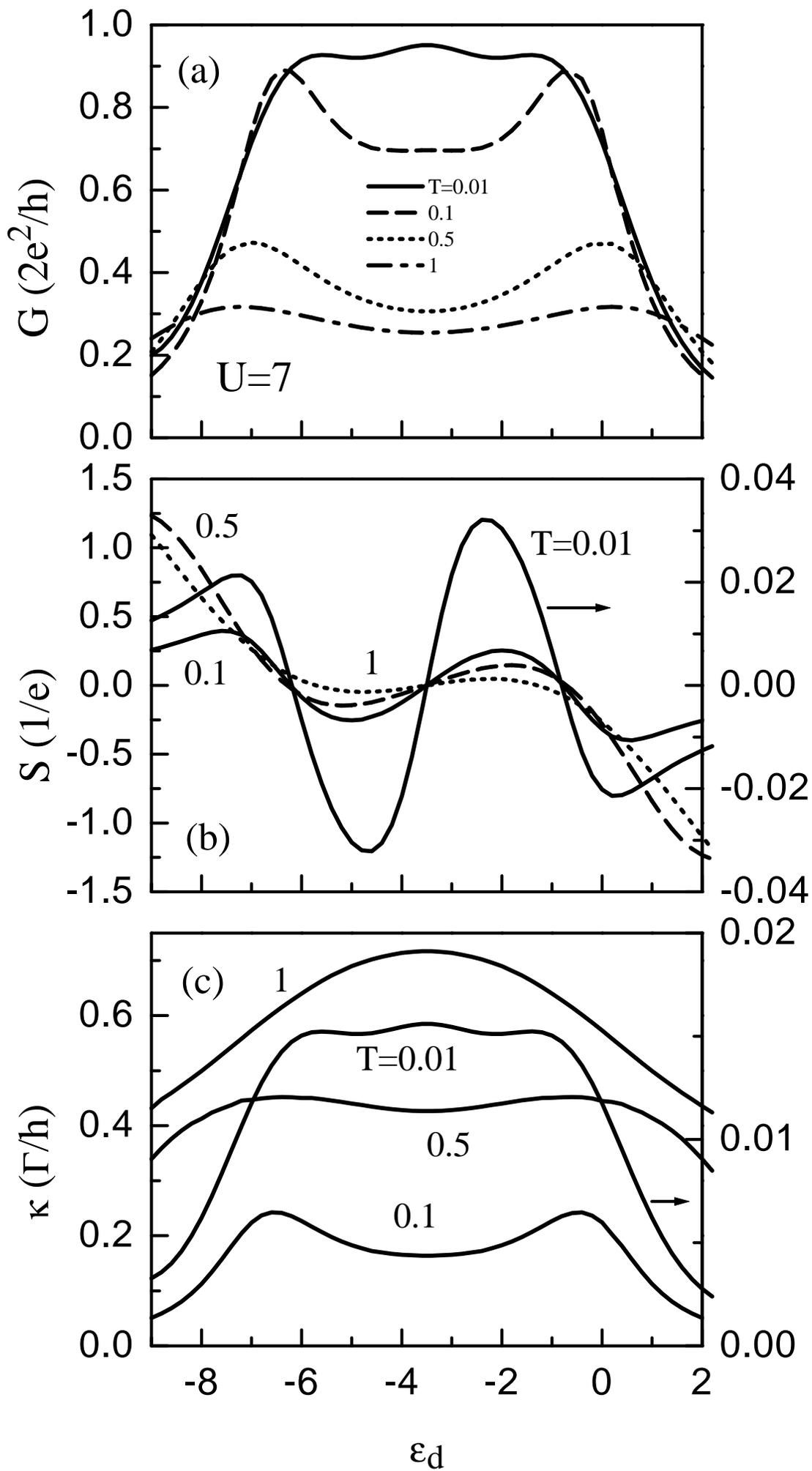
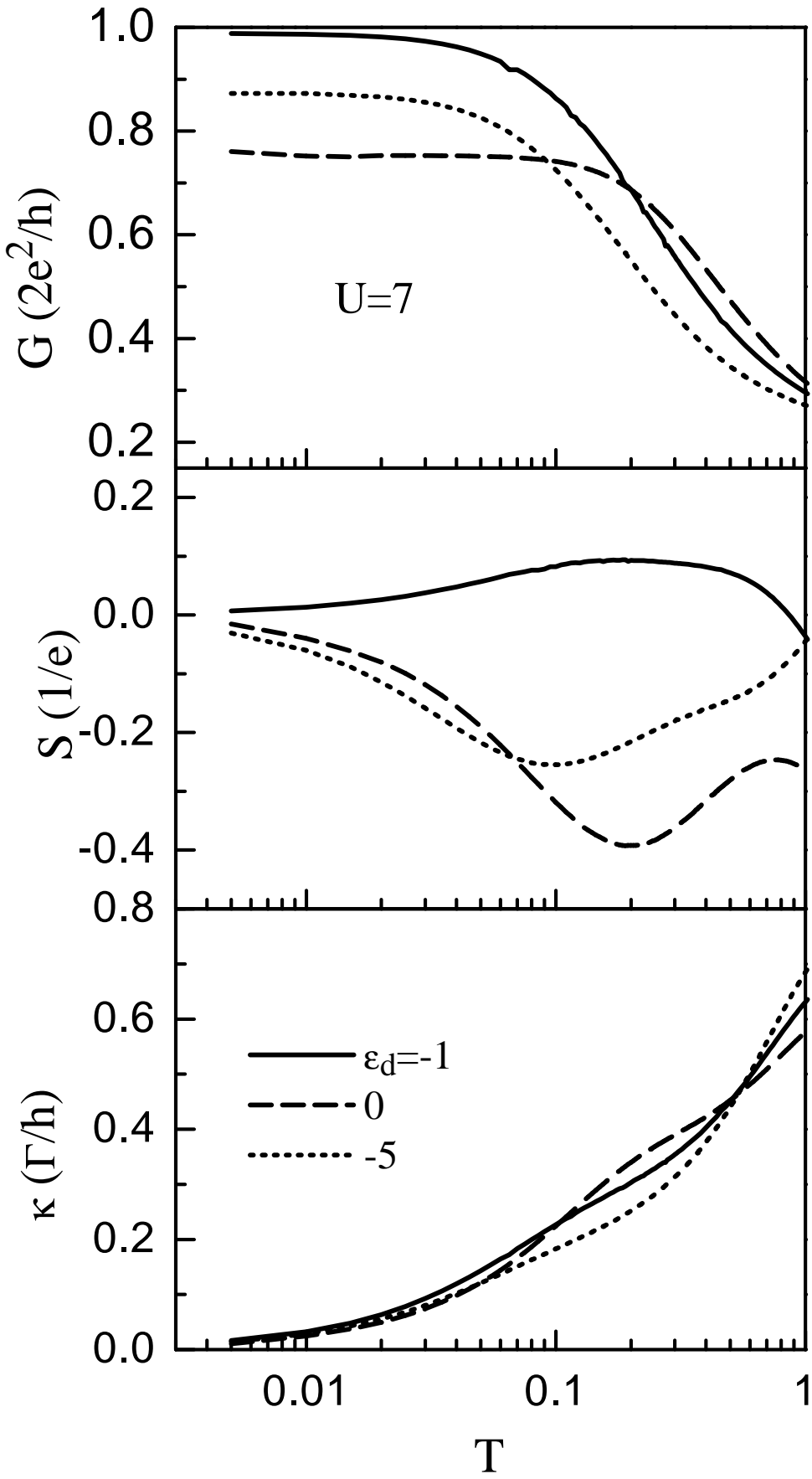
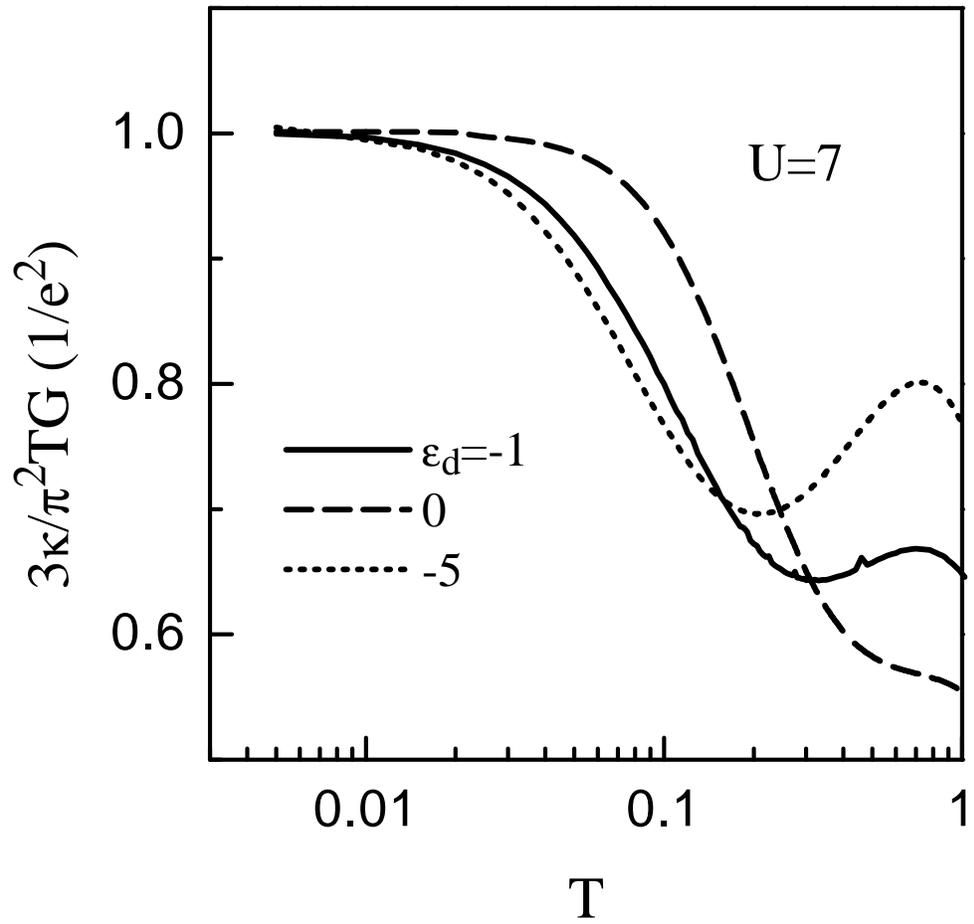
\end{document}